\documentclass[aps, showpacs,pra,superscriptaddress,twocolumn,eqsecnum] {revtex4}
\usepackage{hyperref}
\usepackage{amsmath }
\usepackage{amssymb}
\usepackage{epsfig}
\usepackage{natbib}
\newcommand*{\be}{\begin{equation}}
\newcommand*{\ee}{\end{equation}}

\begin{document}
\bibliographystyle{revtex}
\title{Full-time dynamics of modulational instability in spinor
Bose--Einstein condensates}

\author{Evgeny V. Doktorov}
\email{doktorov@dragon.bas-net.by}

\affiliation{B.I. Stepanov Institute of Physics, 68 Francisk
Skaryna Avenue, 220072 Minsk, Belarus}

\author{Vassilis M. Rothos}

\affiliation{Department of Mathematics, Physics and Computational
Sciences, Faculty of Technology, Aristotle University of
Thessaloniki, Thessaloniki 54124, Greece}

\author{Yuri S. Kivshar}

\affiliation{Nonlinear Physics Centre and Australian Centre of
Excellence for Quantum-Atom Optics, Research School of Physical
Sciences and Engineering, Australian National University, Canberra
ACT 0200, Australia}

\begin{abstract}
We describe the full-time dynamics of modulational instability in
$F=1$ spinor Bose--Einstein condensates for the case of the
integrable three-component model associated with the matrix
nonlinear Schr\"odinger equation. We obtain an exact homoclinic
solution of this model by employing the dressing method which we
generalize to the case of the higher-rank projectors. This
homoclinic solution describes the development of modulational
instability beyond the linear regime, and we show that the
modulational instability demonstrates the reversal property when the
growth of the modulated amplitude is changed by its exponential
decay.
\end{abstract}


\maketitle

\section{Introduction}

Spinor Bose--Einstein condensate (BEC) with an optical confinement
represents a unique macroscopic system with the spin degrees of
freedom~\cite{Exp1,Exp2}. The interplay between the mean-field
effective nonlinearities of three-component matter waves and their
spin properties produce many interesting phenomena such as the
spin mixing~\cite{Exp2}, as well as the formation of spin
domains~\cite{Ketterle1,domain} and spin
textures~\cite{Ohmi,Ketterle2}. Various properties of the spinor
BEC have been analyzed theoretically~\cite{Ho,Pu,Ueda}. The ground
state of the spinor BEC with the hyperfine spin $F=1$ can be
either ferromagnetic (maximum spin projection) or polar (zero spin
projection). It was shown in Ref.~\cite{Robins} within the linear
stability analysis of the spinor condensate model that the
ferromagnetic phase of the condensate can experience instability
for large enough densities of atoms, while the polar phase remains
always modulationally stable.

Wadati and co-authors~\cite{Wadati1} demonstrated that the
three-component nonlinear equations describing the evolution of
the $F=1$  BEC can be reduced, under special constraints imposed
on the condensate parameters, to the completely integrable matrix
nonlinear Schr\"odinger (NLS) equation~\cite{Tsuchida}. Both
bright and dark three-component BEC solitons have been found in
the framework of this
model~\cite{Wadati2,Malomed,Wadati2a,Wadati3,Wadati4}.

As regards the linear stability analysis presented in
Ref.~\cite{Robins}, only an initial (linear) stage of the
perturbation development can be explored by this method which
predicts the exponential growth of the modulation frequency
sidebands for some conditions, i.e., it describes the conditions
of modulational instability (MI). A physical mechanism behind the
MI is the parametric coupling between the spin degrees of freedom
which leads to a population transfer between the spin components.
To study the long-time evolution of instabilities, numerical
methods are used as a rule. For the scalar NLS equation, the
problem of the long-time evolution of the MI was studied by the
truncation of the original model to a finite number of modes (as
usually, the three-mode approximation)~\cite{Trillo}. More
complete analysis of the long-time MI dynamics~\cite{Nail1,Nail2}
is based on a linear constraint imposed on the real and imaginary
parts of solutions of the scalar NLS equation, and it allows one
to find a class of three-parameter solutions sharing this
property. Among the solutions found in such a way, a special
solution describes the development of MI beyond the linear regime,
and it is identified as a homoclinic orbit separating two
qualitatively different types of periodic solutions. A similar
result was obtained by means of the Darboux transformation with
the plane wave as a `seed' solution~\cite{Its}. Following
terminology of Ref. \cite{JNS}, the full-time dynamics represents
a \textit{nonlocal} view of the MI development over a long time
interval.

A homoclinic orbit is a trajectory of a dynamical system that
tends to the same manifold (fixed point, periodic orbit, etc.) as
time tends to $\pm\infty$. The existence of homoclinic solutions
serves as an indicator of chaotic regimes in a perturbed
deterministic system. For nonlinear wave systems described by
partial differential equations, the complete understanding of the
homoclinic structures in the infinite-dimensional phase space is
far from being available at present. On the other hand, the unique
features of the integrable nonlinear wave equations admit
essentially more deep insight into this problem. Extended reviews
of analytical and numerical methods for obtaining homoclinic
orbits for the scalar NLS and sine-Gordon equations are given in
Refs.~\cite{Mac,Mark}.

The aim of our paper is twofold. First, we derive a homoclinic
solution of the matrix NLS equation. Second, using these analytic
results, we present the exact solution of the problem of the
long-time evolution of the modulationally unstable $F=1$ BEC in
the case when it is described by the integrable model.

To find homoclinic solutions, we do not impose {\it ad hoc}
constraints on the form of solutions. Instead, we use a kind of
dressing procedure, well known in the soliton theory~\cite{NMPZ},
which was proposed recently as a systematic tool to generate exact
homoclinic solutions of integrable nonlinear equations with
periodic boundaries \cite{D}. A dressing factor being the main
technical ingredient in this approach contains a projector which
determines the coordinate dependence of the homoclinic solution.
It should be pointed out that for all known homoclinic solutions
obtained up to now for various nonlinear equations (see, e.g.,
Refs.~\cite{D,Wright}), this projector has rank 1. A crucial
feature of the matrix NLS equation consists in the fact that the
corresponding dressing factor incorporates the rank 2 projector.
In terms of the soliton theory it corresponds to \textit{multiple}
zeros of the scattering matrix coefficients (or multiple zeros of
the associated Riemann--Hilbert problem). Notice that the case of
multiple zeros cannot be treated as a coalescence of simple
zeros~\cite{Val}. Accordingly, we modify the definition of the
dressing factor for the case of the matrix NLS equation and obtain
the first example of the \textit{matrix} homoclinic orbit and, as
a result, the complete description of the MI evolution in the
integrable spinor BEC model.

The paper is organized as follows. In Sec. II we describe the
integrable $F=1$ BEC model. The method for obtaining homoclinic
solutions for integrable nonlinear equations valid for higher rank
projectors is outlined in Sec. III. Section IV is devoted to the
explicit derivation of the homoclinic solution for the matrix NLS
equation and presents the main results of our paper. The
homoclinic solution describes the temporal evolution of
\textit{linearly unstable} modes. We show that the MI has a
reversal property -- the initial-wave profile is recovered after a
sufficiently long time. Hence, the term 'side-band instability'
refers in fact to only the linear stage of the instability
development. Section V concludes the paper.

\section{Model}

We consider an effective one-dimensional BEC trapped in a
pencil-shaped region elongated in the $x$ direction and tightly
confined in the transversal directions. The assembly of atoms in the
hyperfine spin $F=1$ state is described by a vector order parameter
$\vec{\Phi}(x,t)=(\Phi_+(x,t),\Phi_0(x,t),\Phi_-(x,t))^T$, where its
components correspond to three values of the spin projection
$m_F=1,0,-1$. The functions $\Phi_\pm$ and $\Phi_0$ obey a system of
coupled Gross--Pitaevskii equations \cite{Suomi,Wadati2}
\begin{eqnarray}\label{sys1}
i\hbar\partial_t\Phi_\pm=&-&\frac{\hbar^2}{2m}\partial_x^2\Phi_\pm+(c_0+c_2)(|\Phi_\pm|^2
+|\Phi_0|^2)\Phi_\pm\nonumber\\
&+&(c_0-c_2)|\Phi_\mp|^2\Phi_\pm+c_2\Phi_\mp^*\Phi_0^2,\\
i\hbar\partial_t\Phi_0=&-&\frac{\hbar^2}{2m}\partial_x^2\Phi_0+(c_0+c_2)(|\Phi_+|^2+|\Phi_-|^2)\Phi_0
\nonumber\\
&+&c_0|\Phi_0|^2\Phi_0+2c_2\Phi_+\Phi_-\Phi_0^*,\nonumber
\end{eqnarray}
where the constant parameters $c_0=(g_0+2g_2)/3$ and
$c_2=(g_2-g_0)/3$ control the spin-independent and spin-dependent
interaction, respectively. The coupling constant $g_f$ ($f=0,2$)
is given in terms of the $s$-wave scattering length $a_f$ in the
channel with the total hyperfine spin $f$,
\[
g_f=\frac{4\hbar^2a_f}{ma_\perp^2}\left(1-C\frac{a_f}{a_\perp}\right)^{-1}.
\]
Here $a_\perp$ is the size of the transverse ground state, $m$ is
the atom mass, and $C=-\zeta(1/2)\approx1.46$.

 It was noted in \cite{Wadati1} that Eqs. (\ref{sys1}) are
 reduced to an integrable system under the constraint
 \be\label{const}
 c_0=c_2\equiv-c<0.
 \ee
 The negative $c_2$ means that we consider the ferromagnetic ground state of the
 spinor BEC with attractive interactions.
 The condition (\ref{const}), being written in terms of $g_f$ as
 $2g_0=-g_2>0$,
 imposes a constraint on the scattering lengths:
 $a_\perp=3Ca_0a_2/(2a_0+a_2)$. Redefining the function
 $\vec{\Phi}$ as
 $\vec{\Phi}\rightarrow(\phi_+,\sqrt{2}\phi_0,\phi_-)^T$,
 normalizing the coordinates as $t\rightarrow(c/\hbar)t$ and
 $x\rightarrow(\sqrt{2mc}/\hbar)x$, and accounting for the
 constraint (\ref{const}), we obtain a reduced system of equations
 in a dimensionless form:
 \be
 i\partial_t\phi_\pm+\partial_{x}^2\phi_\pm+2\left(|\phi_\pm|^2+2|\phi_0|^2\right)\phi_\pm+2\phi_
 \mp^*\phi_0^2=0\;,\label{sys2}
 \ee
 \[
 i\partial_t\phi_0+\partial_{x}^2\phi_0+2\left(|\phi_+|^2+|\phi_0|^2+|\phi_-|^2\right)\phi_0
 +2\phi_+\phi_0^*\phi_-=0\;.
 \]
 After arranging the components $\phi_\pm$ and $\phi_0$ into a
 $2\times2$ matrix $Q=\left(\begin{array}{cc}\phi_+ & \phi_0\\
 \phi_0& \phi_-\end{array}\right)$, we can easily see that
 Eqs. (\ref{sys2}) take the form of the integrable matrix NLS
 equation
 \be
 i\partial_tQ+\partial_x^2Q+2QQ^\dag Q=0\;.\label{MNLS}
 \ee
 The matrix NLS equation (\ref{MNLS}) possesses the Lax
 representation with the $4\times4$ matrices $U$ and $V$ of the
 form \cite{Tsuchida}
 \begin{gather}
 U=ik\Lambda+\hat Q, \; \Lambda=\mathrm{diag}(-1,-1,1,1), \;
 \hat Q=\left(\begin{array}{cc}0 & Q\\
 -Q^\dag & 0\end{array}\right)\;, \label{U} \\
 V=2ik^2\Lambda+2k\hat Q+i\left(\begin{array}{cc} QQ^\dag & Q_x\\
 Q^\dag_x & -Q^\dag Q\end{array} \right)\;,\label{V}
 \end{gather}
 $k$ is a spectral parameter.

 \section{Method}

 We are interested in periodic solutions of Eqs. (\ref{sys2})
(or (\ref{MNLS})) with a spatial period $L$, $Q(x+L,t)=Q(x,t)$.
 Hence, the Floquet theory should be applied to analyze the
 spectral problem
 \begin{equation}
 M_x=UM\;. \label{SP}
 \end{equation}
 The fundamental solution $M(x,k)$ of Eq. (\ref{SP}) is fixed by the
 condition $M(0,k)=I$, $I$ is the unit $4\times4$ matrix. Then we
 define a transfer matrix $T(k)$ as the fundamental solution in the
 point $x=L$, $T(k)=M(L,k)$. Diagonalization of the transfer
matrix determines a matrix $R$,
\[
R^{-1}T(k)R=\mathrm{diag}\left(e^{im_1L}, \ldots,
e^{im_4L}\right)\equiv \Delta(L,k),
\]
and produces the Floquet multipliers $\exp(im_jL)$ with the
Floquet exponents $m_j$, $j=1,\ldots,4$. The Floquet spectrum is a
set of all $k$ for which the transfer matrix $T(k)$ has the
eigenvalues on the unit circle.

The next step is a determination of a Bloch solution $\chi$ of Eq.
(\ref{SP}) as $\chi=MR$ which obeys the property
$\chi(x+L,k)=\chi(x,k)\Delta(L,k)$, specific for the Bloch-type
solutions. The Bloch eigenfunctions of the periodic spectral
problem (\ref{SP}) play the role of the Jost solutions of the
spectral problem with a decreasing potential.

Among the points of the Floquet spectrum we will distinguish the
so called double points \cite{JNS}. Double points are those values
of $k$ for which the Floquet exponents $m_j$ differ in multipliers
of $2\pi/L$ or, in other words, the Floquet multipliers are
degenerate. We will be especially interested in complex double
points which indicate linearized instability of solutions of Eq.
(\ref{sys2}) and label orbits homoclinic to hyperbolic fixed
points in the phase space of a nonlinear system. Note that the
term ``double" in the context of the Floquet spectrum refers to
the algebraic multiplicity of a point of the spectrum and has no
relation to the multiplicity of zeros we have spoken about in the
Introduction. Real double points are associated with stable modes.

 Suppose we know explicitly a Bloch solution $\chi_0$ of the
 spectral problem $\chi_{0x}=U_0\chi_0$ with the matrix $U_0$ (\ref{U}) whose
 entries contain the known solution $Q_0$ of Eq. (\ref{MNLS}). Then we
 dress the solution $\chi_0$ by applying the dressing factor
 $D(x,t,k)$, $\chi=D\chi_0$, and $\chi$ is a new solution of the
 spectral problem with new matrix $U=DU_0D^{-1}+D_xD^{-1}$. The
 dressing factor has the form
 \be\label{DF}
 D=I-\sum_{s=1}^{N}\frac{k_s-k_s^*}{k-k_s^*}P_s(x,t),
 \ee
 where $P_s$ is a projector, $P_s^2=P_s$,
 \[
 P_s=\frac{1}{k_s-k_s^*}\sum_{n,l=1}^{r_s}|n;s\rangle(D^{(s)-1})_{nl}\langle
 l;s|,
 \]
 \be\label{proj}
 D_{nl}^{(s)}=\frac{\langle
 n;s|l;s\rangle}{k_s-k_s^*}.
 \ee
 Here $k_s$, $s=1,\ldots,N$ are complex double points of the Floquet spectrum and
 $r_s$ is the rank of the projector $P_s$. The
 four-component ket- and bra-vectors $|n;s\rangle$
 and $\langle l;s|$ are the column and row arrays,
 respectively. Hence,
 $|n;s\rangle$
 is a four vector related with the $s$th complex double
 point $k_s$ and obtained by applying
 the Bloch solution $\chi_0(k_s)$ to a
 constant vector $|q;s\rangle$,
 \be\label{vec}
 |n;s\rangle=\chi_0(x,t,k_s)|q;s\rangle.
 \ee
 Exactly $r^{(s)}$ vectors $|q;s\rangle$, and hence $r^{(s)}$ vectors
 $|n;s\rangle$, correspond to the complex double point $k_s$.
 The summation in Eq. (\ref{DF}) is taken over all $N$ complex
 double points, while that in Eq. (\ref{proj}) is performed over
 the $r_s$-dimensional space of vectors $|n;s\rangle$ produced
 in accordance with Eq. (\ref{vec}).
 Then a new solution of the matrix NLS equation is written as
 \be\label{new}
 \hat Q=\hat Q_0+\sum_{s=1}^{N}(k_s-k_s^*)[\Lambda,P_s].
 \ee
 For the rank 1 projectors these formulas reduce to the known ones
 \cite{D}.

 Note the essential difference in applications of the dressing
 procedure between the soliton theory and the periodic wave
 theory.
 Indeed, the parameters $k_s$ are free in the standard use
 of the dressing method and, in any case, they do note relate with
 the seed solution $\hat Q_0$. On the contrary, our approach demands to choose
 $k_s$ as the complex double points of the Floquet spectrum of the
 spectral problem (\ref{SP}) for the seed solution $\chi_0$.

 Hence, they are the complex double points $k_s$ and the
 projectors
 $P_s$ that completely determine new solution. In the next section
 the above method will be used to generate homoclinic solution of
 the spin 1 BEC model (\ref{sys2}) and hence to reveal the
 long-time dynamics of the MI in this model.

 \section{Results}

 We begin with a spatially homogeneous continuous wave solution of
 Eq.
 (\ref{sys2}) with components
 \be\label{wave}
 \phi_+^{(0)}=\phi_-^{(0)}=ae^{-i\mu t}, \quad
 \phi_0^{(0)}=ibe^{-i\mu t}
 \ee
 as the seed solution to be dressed.  Here $a$ and $b$ are
 real constant amplitudes which determine a population of each
 spin component, and the chemical potential $\mu$ is given by
 $\mu=-2(a^2+b^2)$. Note the fixed $\pi/2$ phase difference
 between the components $\phi_\pm^{(0)}$ and $\phi_0^{(0)}$. The same phase
 locking property is an inherent feature of the nonintegrable model
 (\ref{sys1}) as well \cite{Robins}.
  We could start with
  more general representation of plane waves but the structure of
  Eqs. (\ref{sys2}) and the Galilean invariance reduce it to
  the form (\ref{wave}). Then we consider the spectral problem
 (\ref{SP}) with the matrix $U_0$ containing the plane waves
 (\ref{wave}) as the potential $Q_0$:
 \be\label{U_0}
 U_0=\left(\begin{array}{cccc}-ik & 0 & ae^{-i\mu t} & ibe^{-i\mu
 t}\\ 0& -ik & ibe^{-i\mu t} & ae^{-i\mu t}\\ -ae^{i\mu t} & ibe^{i\mu
 t} & ik & 0\\ ibe^{i\mu t} & -a e^{i\mu t} & 0&
 ik\end{array}\right).
 \ee

 The fundamental solution of the spectral problem with the matrix
 $U_0$ is explicitly found:
 \begin{widetext}
 \be\label{M}
 M=\left(\begin{array}{cccc} \cos px+i(k/p)\sin px & 0 &
 (a/p)\sin px e^{-i\mu t} & i(b/p)\sin px e^{-i\mu t}
 \\ 0& \cos px+i(k/p)\sin px  & i(b/p)\sin px e^{-i\mu
 t} & (a/p)\sin px e^{-i\mu t} \\ -(a/p)\sin px e^{i\mu
 t} & i(b/p)\sin px e^{i\mu t} & \cos px-i(k/p)\sin px &
 0\\ i(b/p)\sin px e^{i\mu t} & -(a/p)\sin px e^{i\mu
 t} & 0 & \cos px-i(k/p)\sin px \end{array}\right), \quad
 \det M=1,
 \ee
 \end{widetext}
 where $p^2=a^2+b^2+k^2$.
 Diagonalization of the transfer matrix
 $T(k)=M(L,k)$ is performed by the matrix $R$ which has the form
 \begin{widetext}
 \be\label{R}
 R=\left(\begin{array}{cccc} d_1 & -i(a/b)d_2 &
 [b/(p+k)]d_3e^{-i\mu t} & -i[a/(p+k)]d_4e^{-i\mu t} \\ -i(a/b)d_1 &
 d_2 & -i[a/(p+k)]d_3e^{-i\mu t} & [b/(p+k)]d_4e^{-i\mu t}
 \\ 0& -[(p-k)/b]d_2e^{i\mu t} & 0 & d_4\\ -[(p-k)/b]d_1e^{i\mu t}
 & 0 & d_3 & 0 \end{array}\right),
 \ee
 \end{widetext}
 where $d_j$ are time dependent. As a result,
 \[
 R^{-1}T(k)R=\Delta(L,k)=\mathrm{diag}(e^{-ipL}, e^{-ipL}, e^{ipL},
 e^{ipL})\;.
 \]
 Therefore, the Floquet exponents are written as
 \begin{equation}\label{exp}
 m_1=-p, \quad m_2=-p, \quad m_3=p, \quad m_4=p
 \end{equation}
 and have the multiplicity 2.
 Then we obtain the seed Bloch solution $\chi_0=MR$ in the form
 \[
 \chi_0=\exp(i\frac{\mu}{2}t)\!\!\left(\begin{array}{cccc} d_{10} &
 -i\frac{a}{b}d_{20} & \frac{b}{p+k}d_{30} & -\frac{ia}{p+k}d_{40}
 \\ -i\frac{a}{b}d_{10} & d_{20} & -\frac{ia}{p+k}d_{30} &
 \frac{b}{p+k}d_{40}\\ 0 & \frac{k-p}{b}d_{20} & 0 & d_{40} \\
 \frac{k-p}{b}d_{10} & 0 & d_{30} & 0 \end{array}\right)
 \]
 \be
 \times \exp(ip\Lambda x+2ikp\Lambda t),\label{Bloch}
 \ee
 where the parameters $d_j(t)$ entering Eq. (\ref{R}) have been
 determined from the second Lax equation $\chi_{0t}=V_0\chi_0$ with the
 matrix $V_0$ (\ref{V}) depending on the seed continuous wave
 (\ref{wave}):
 \[
 d_1=d_{10}\exp(-2ik^2t), \quad d_2=d_{20}\exp(2ik^2t),
 \]
 \[
  d_3\!=\!d_{30}\exp\!\left[-\frac{i}{2}\mu t-2ikpt\right], \;
  d_4\!=\!d_{40}\exp\!\left[\frac{i}{2}\mu t+2ikpt\right].
  \]
  Here $d_{j0}$ are integration constants.

  Now we proceed to finding the complex double points. Following
  Ref.
 \cite{JNS}, we seek for double points as a difference between two
 Floquet exponents $m_1$ and $m_3$ (\ref{exp}):
 \[
 m_3=m_1+\delta_s, \qquad \delta_s=\frac{2\pi}{L}s, \quad
 s=\pm1,\pm2, \ldots\;.
 \]
 This gives
 \be\label{complex}
 k_s=\pm i\sqrt{a^2+b^2-(\pi s/L)^2}, \quad \mathrm{if} \quad
 a^2+b^2>(\pi s/L)^2
 \ee
 and
 \be\label{real}
 k_s=\pm\sqrt{(\pi s/L)^2-(a^2+b^2)}, \quad \mathrm{if} \quad
 a^2+b^2<(\pi s/L)^2\;.
 \ee
 Hence, for given amplitudes $a$ and $b$ and period $L$ the double
 points  are arranged into infinite number of real double points
 (\ref{real})
 situated on the real axis in the $k$ plane, and a finite number of
 complex double points (\ref{complex}) lying on the imaginary axis within the
 interval $(-i\sqrt{a^2+b^2},i\sqrt{a^2+b^2})$.

 Let us choose in the following the
 amplitudes and period in such a way that to obtain the single
 complex double point $k_1$ (and hence $-k_1$). It means $\sqrt{a^2+b^2}>(\pi/L)$ but
 $\sqrt{a^2+b^2}<(2\pi/L$). For this choice the only rank 2 projector
 $P\equiv P_1$ has the form
 \[
 P=\frac{1}{k_1-k_1^*}\sum_{n,l=1}^{2}|n\rangle(D^{-1})_{nl}\langle
 l|\;, \quad D_{nl}=\frac{\langle n|l\rangle}{k_1-k_1^*}\;.
 \]
 To simplify notations, we write $|n\rangle$ instead of
 $|n;1\rangle$.
 Since $D$ is a $2\times2$ matrix, we easily obtain the following
 expression for the projector:
 \be\label{expl}
 P=\frac{1}{\widetilde D}[\langle 2|2\rangle|1\rangle\langle 1|-\langle 1|2\rangle|1\rangle\langle
 2|-\langle 2|1\rangle|2\rangle\langle 1|+\langle
 1|1\rangle|2\rangle\langle
 2|],
 \ee
 \[
 \widetilde D=\langle 1|1\rangle\langle 2|2\rangle-\langle 1|2\rangle\langle
 2|1\rangle,
 \]
 where the vectors $|1\rangle$ and $|2\rangle$ are determined as
 \[
 |1\rangle=\chi_0(k_1)|q\rangle, \qquad
 |2\rangle=\chi_0(k_1)|r\rangle\;.
 \]
 Here $|q\rangle$ and $|r\rangle$ are linearly independent constant vectors with the components $q_j$ and
 $r_j$, $j=1,\ldots,4$.
 Then
 after rather lengthy but straightforward algebraic calculation in
 accordance with Eqs. (\ref{expl}) and (\ref{new}) taken for $s=1$, we explicitly obtain new solutions of the
 integrable spin 1 BEC model (\ref{sys2}),
 \be\label{newsol}
 \phi_+=\phi_-=ae^{-i\mu t}\left(1+2i\frac{B}{A}\sin\psi\right),
 \ee
 \[
 \quad \phi_0=ibe^{-i\mu t}\left(1+2i\frac{B_0}{A}\sin\psi\right),
 \]
 which at the same time represent components of the matrix
 homoclinic orbit of the matrix NLS equation (\ref{MNLS}). Here
 \be\label{A}
 A=\cosh2\tau-\cos2\rho\sin^2\psi+2\gamma\cosh\tau\sin\rho\sin\psi+\frac{1}{2}\gamma^2,
 \ee
 \[
 B=\sinh2\tau\cos\psi+i\cosh2\tau\sin\psi-i\cos2\rho\sin\psi
 \]
 \[
 +\frac{1}{4}\gamma\left[(\mu_-e^\tau
 +\mu_-^*e^{-\tau})e^{i\rho}-(\mu_+^*e^\tau+\mu_+e^{-\tau})e^{-i\rho}\right]
 \]
 \be\label{B}
 +
 \frac{i}{2}\gamma^2\left(\sin\psi-\frac{b}{a}\cos\psi\cos\alpha_{23}\right),
 \ee
 \[
 B_0=\sinh2\tau\cos\psi+i\cosh2\tau\sin\psi-i\cos2\rho\sin\psi
 \]
 \[
 +\frac{1}{4}\gamma\left[(\nu_-e^\tau
 +\nu_-^*e^{-\tau})e^{i\rho}-(\nu_+^*e^\tau+\nu_+e^{-\tau})e^{-i\rho}\right]
 \]
 \be\label{B0}
 +
 \frac{i}{2}\gamma^2\left(\sin\psi-\frac{a}{b}\cos\psi\cos\alpha_{23}\right).
 \ee
 The constants $d_{j0}$ have been incorporated into the constant
 components $q_j$ and $r_j$ of the
 vectors $|q\rangle$ and $|r\rangle$.
 If we denote definite combinations of these components  as
 \[
 e_1=q_1r_2-q_2r_1, \quad e_2=q_1r_3-q_3r_1, \quad
 e_3=q_1r_4-q_4r_1,
 \]
 \[
 e_4=q_3r_4-q_4r_3, \; e_j=|e_j|e^{i\alpha_j}, \; \alpha_{jl}=\alpha_j-\alpha_l,
 \; |e_2|=|e_3|,
 \]
 then the notations used in (\ref{A}), (\ref{B})
 and (\ref{B0}) are as follows:
 \[
 k_1=ik_0,\quad p_1=\sqrt{a^2+b^2+k_1^2}=\frac{\pi}{L}, \quad
 e^{i\psi}=\frac{p_1+ik_0}{\sqrt{a^2+b^2}},
 \]
 \[
 \tau=4k_0p_1t+t_0, \quad \rho=2p_1x-\alpha_{13}, \quad \alpha_{13}=\alpha_{34},
 \]
 \[
 e^{t_0}=\frac{\sqrt{a^2+b^2}}{b}\sqrt{\frac{|e_1|}{|e_4|}},\quad
 \gamma=\frac{2|e_2|}{\sqrt{|e_1||e_4|}},
 \]
 \[
 \mu_\pm=1\pm ie^{\mp
 i\psi}\left(\sin\psi-\frac{b}{a}e^{i\alpha_{23}}\cos\psi\right),
 \]
 \[
 \nu_\pm=1\pm ie^{\mp
 i\psi}\left(\sin\psi+\frac{a}{b}e^{i\alpha_{23}}\cos\psi\right).
 \]
 The solutions (\ref{newsol}) are indeed homoclinic to the plane
 waves (\ref{wave}). Calculation of the asymptotics of $\phi_\pm$ and
 $\phi_0$ as $t\rightarrow \pm\infty$ gives
 \[
 \phi_\pm \rightarrow ae^{-i\mu t}e^{\pm2i\psi}, \quad
 \phi_0 \rightarrow ibe^{-i\mu t}e^{\pm2i\psi}.
 \]
 In other words, these solutions reproduce in the limit
 $t\rightarrow\pm\infty$ the seed plane waves up to a constant
 phase, as should be for the homoclinic orbit. Note that the
 nonlinear MI for the spin 1 condensate but for different phases
 of the seed wave components was studied by the Darboux
 transformation in Ref. \cite{Malomed}.

\begin{figure}
 \begin{center}
 \includegraphics[scale=0.6]{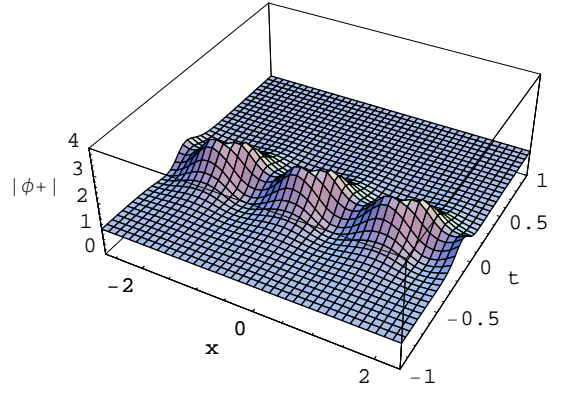}
 \caption{Full-time evolution of the $\phi_+$ (and $\phi_-$) component due to modulational instability.
 The parameters
 are $a=1$, $b=2$, $L=\pi/2$, $\alpha_2=\pi/3$, $\alpha_{3}=\pi/4$, $|e_j|=1$.}
 \label{Figure1}
 \end{center}
 \end{figure}

\begin{figure}
 \begin{center}
 \includegraphics[scale=0.6]{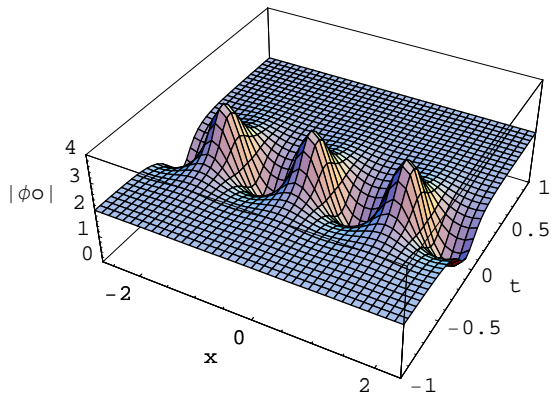}
 \caption{Full-time evolution of the $\phi_0$ component due to modulational instability. The parameters
 are the same as in Fig.~\ref{Figure1}.}
 \label{Figure2}
 \end{center}
 \end{figure}

 Figures \ref{Figure1} and \ref{Figure2} illustrating the solution
 (\ref{newsol})
 demonstrate typical development of
 the continuous wave perturbation within three periods in $x$.
 We see that the stage of the
 exponential growth of instabilities revealed by the linear
 stability analysis transforms to the exponential decreasing with
 emergence of localized structures. Hence, the full-time evolution of MI for the
 integrable $F=1$ BEC
 model demonstrates the reversal property, such as the
 Fermi--Pasta--Ulam process \cite{AC}: the phase trajectory of the system
 returns to the initial one which corresponds to the continuous waves (\ref{wave}).
 For chosen parameters the growth
 and decrease development of the component $\phi_0$ is more  pronounced
 than that of $\phi_\pm$.

 \section{Conclusions}

 We have derived the analytic formulas for describing the full-time
 dynamics of the modulational instability in the integrable model
 of $F=1$ Bose--Einstein condensates. Our results are based on
 the exact homoclinic solution of the matrix NLS equation with the
 continuous plane waves as an initial condition. We have shown that
 there exist cycles of the MI evolution with the reversal property when the
 exponential growth of the modulation amplitude changes to its exponential
 decay. The solution we present here is an example of \textit{large-amplitude} periodic solutions.
 It describes an exponential growth of a weak modulation
 of a background for an initial stage of the condensate evolution,
 and in this sense the background is unstable. However, in the
 nonlinear regime this exponentially growing mode saturates and
 subsequently transforms into oscillations. As expected, the
 integrable model (\ref{sys2}) does
 not exhibit long-time chaotic dynamics contrary to
 the regimes observed numerically for a general case~\cite{Robins},
 but it may serve as a good analytical approximation of the evolution
 of the condensate experiencing the instability. Higher-order homoclinic
 solutions which correspond to several complex double points can be obtained
 by the method described in~\cite{D}.

 Strictly speaking, the analysis based on the continuous wave
 model is not applicable to the trapped systems. Nevertheless,
 such an approach remains valid when the typical spatial
 extent of the condensate is larger than the period of the localized
 pattern formed in result of the instability. More realistic models should
 account for a (small) deviation of the condensate parameters from the constraint
 which provides integrability of the model.
 There exists an approach~\cite{Li} to reveal a persistence of the
 homoclinic orbit when the integrability condition breaks, and therefore to
 establish analytically the existence of chaos. This approach is based on the construction
of the so-called Melnikov function from the quadratic products of
the Bloch functions evaluated on the homoclinic orbit. In this
paper we have explicitly built all the ingredients to perform the
Melnikov analysis. Corresponding results will be published
elsewhere.

\end{document}